\documentclass[american,prl,preprint]{revtex4}
\usepackage{babel}
\usepackage{graphics}

\makeatletter

\usepackage{bm}
\def\Bias{\mathop{\operator@font Bias}\nolimits}
\def\E{\mathop{\operator@font E}\nolimits}
\def\P{\mathop{\operator@font P}\nolimits}

\makeatother
\begin{document}

\title{Stochastic View of Photon Migration in Turbid Media}

\author{M. Xu}

\author{W. Cai}

\author{M. Lax}

\author{R. R. Alfano}

\address{Institute for Ultrafast Spectroscopy and Lasers, New York State Center of
Advanced Technology for Ultrafast Photonics, and Department of Physics,
The City College and Graduate Center of City University of New York,
New York, NY 10031}

\email{minxu@sci.ccny.cuny.edu}

\begin{abstract}
Light propagation in an infinite uniform turbid medium is treated
as a Markov stochastic process of photons to provide an intuitive
framework for photon migration. The macroscopic physical quantities
of photon migration are shown to be completely determined by the microscopic
statistics of the photon propagation direction in direction space
and a generalized Poisson distribution of the number of scattering
events that includes an exponential decay absorption factor. A proper
diffusion solution is derived with an exact time-dependent central
position and half width of photon migration in this framework. The
diffusion coefficient is found to be absorption-independent.
\end{abstract}
\maketitle
Propagation in a multiple scattering (turbid) medium such as the atmosphere
is commonly treated by the theory of radiative transfer (see, for
example, Chandrasekhar's classic text\cite{		  chandrasekhar50:_radia}).
Recent advances in ultrafast lasers and photon detectors for biomedical
imaging and diagnostics have revitalized this field\cite{	  alfano94:_advan,	  yodh97:_diffus,	  gandjbakhche99:_proceed_inter_instit_works_optic_imagin_nih}.
The basic equation of radiative transfer is the elastic Boltzmann
equation, a non-separable integro-differential equation of first order
for which an exact closed form solution is not known except for the
case for isotropic scatterers as far as the authors know\cite{	  hauge74:_tba}.
Solutions are often based on truncation of the spherical harmonics
expansion of the photon distribution function or resort to numerical
calculation including Monte Carlo simulation\cite{		  ishimaru78:_wave,		  cercignani88:_boltz}.
Cai et. al.\cite{	  cai00:_cumul_boltz} recently used truncation
of a cumulant expansion of the photon distribution function to solve
the elastic Boltzmann equation.

In this Letter, photon migration in an infinite uniform medium is
treated as a Markov stochastic process. The solution to the elastic
Boltzmann equation with a point source propagating initially along
the positive \( z \)-axis from the origin of space and time is interpreted
as the probability of finding a photon at any specified location,
direction and time.

In the stochastic picture of photon migration in turbid media, photons
take a random walk in the medium and may get scattered or absorbed
according to the scattering coefficient \( \mu _{s} \) and the absorption
coefficient \( \mu _{a} \) of the medium. A normalized phase function,
\( f(\mathbf{s}\cdot \mathbf{s}') \), describes the probability of
scattering a photon from direction \( \mathbf{s} \) to \( \mathbf{s}' \).
The free path (step-size) between consecutive events (either scattering
or absorbing) has an exponential distribution \( \mu _{T}\exp (-\mu _{T}d) \)
characterized by the total attenuation \( \mu _{T}=\mu _{s}+\mu _{a} \).
At an event, photon scattering takes place with probability \( \mu _{s}/\mu _{T} \)
(the albedo) and absorption with probability \( \mu _{a}/\mu _{T} \).
This picture forms the basis for Monte Carlo simulation of photon
migration.

Here we will show that this simple picture of a Markov stochastic
process of photons can be utilized to compute analytically macroscopic
quantities such as the average central position and half width of
the photon distribution. The microscopic statistics of the photon
propagation direction in direction space (solely determined by the
phase function and the incident direction) provides a basis for computing
the macroscopic quantities at any specified time and position. The
bridge between them is a generalized Poisson distribution \( p_{n}(t) \),
the probability that a photon has endured exactly \( n \) scattering
events before time \( t \) (solely determined by the scattering and
absorption coefficients of the medium).

Denote the \( i \)th propagation position, direction and step-size
of a photon as \( \mathbf{x}^{(i)} \), \( \mathbf{s}^{(i)} \) and
\( d^{(i)} \). The initial condition is \( \mathbf{x}^{(0)}=(0,0,0) \)
for the starting point and \( \mathbf{s}^{(0)}=\mathbf{s}_{0}=(0,0,1) \)
for the incident direction. The laboratory Cartesian components of
\( \mathbf{x}^{(i)} \) and \( \mathbf{s}^{(i)} \) are \( x^{(i)}_{\alpha } \)
and \( s^{(i)}_{\alpha } \) (\( \alpha =1,2,3 \)). The photon is
incident at time \( t_{0}=0 \). For simplicity the speed of light
is taken as the unit of speed and the mean free path \( \mu _{T}^{-1} \)
as the unit of length.

\begin{figure}
{\centering \includegraphics{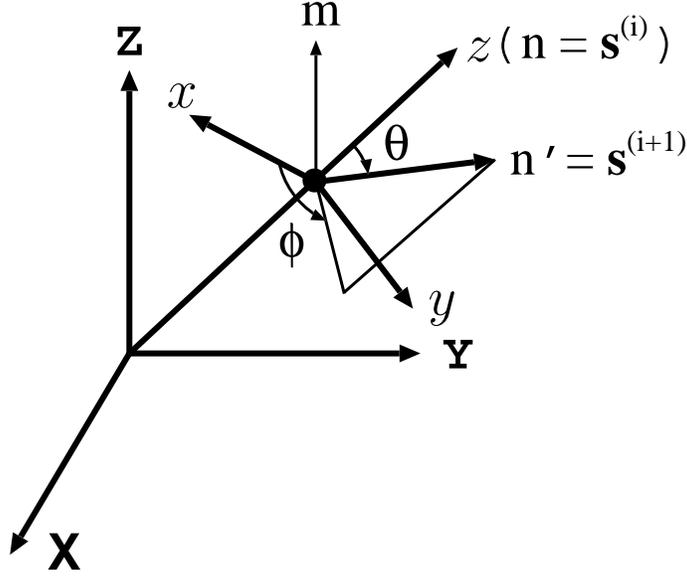} \par}

\caption{A photon moving along \protect\( \mathbf{n}\protect \) is scattered
to \protect\( \mathbf{n}'\protect \) with a scattering angle \protect\( \theta \protect \)
and an azimuthal angle \protect\( \phi \protect \) in a photon coordinate
system \protect\( xyz\protect \) whose \protect\( z\protect \)-axis
coincides with the photon's propagation direction prior to scattering.
\protect\( XYZ\protect \) is the laboratory coordinate system.}

\label{fig:photon-scattering}
\end{figure}

The scattering of photons takes a simple form in an orthonormal coordinate
system \( (\frac{\mathbf{m}-(\mathbf{n}\cdot \mathbf{m})\mathbf{n}}{\left| \mathbf{n}\times \mathbf{m}\right| },\frac{\mathbf{n}\times \mathbf{m}}{\left| \mathbf{n}\times \mathbf{m}\right| },\mathbf{n}) \)
attached to the moving photon itself where \( \mathbf{n} \) is the
photon's propagation direction prior to scattering and \( \mathbf{m} \)
is an arbitrary unit vector not parallel to \( \mathbf{n} \){[}see
Fig.~\ref{fig:photon-scattering}{]}. The distribution of scattering
angle \( \theta \in [0,\pi ] \) is given by the phase function of
the medium and the azimuthal angle \( \phi	 \) is uniformly distributed
over \( [0,2\pi ) \). For one realization of the scattering event
of angles \( (\theta ,\phi ) \) in the photon coordinate system,
the outgoing propagation direction \( \mathbf{n}' \) of the photon
will be:\begin{equation}
\label{eq:1}
\mathbf{n}'=\frac{\mathbf{m}-(\mathbf{n}\cdot \mathbf{m})\mathbf{n}}{\left| \mathbf{n}\times \mathbf{m}\right| }\sin \theta \cos \phi +\frac{\mathbf{n}\times \mathbf{m}}{\left| \mathbf{n}\times \mathbf{m}\right| }\sin \theta \sin \phi +\mathbf{n}\cos \theta .
\end{equation}
The freedom of choice of the unit vector \( \mathbf{m} \) reflects
the arbitrariness of the \( xy \) axes of the photon coordinate system.
For example, taking \( \mathbf{m}=(0,0,1) \), Eq.~(\ref{eq:1})
gives\begin{eqnarray}
s^{(i+1)}_{1} & = & -\frac{\sin \theta }{\sqrt{1-\left( s^{(i)}_{3}\right) ^{2}}}(s^{(i)}_{1}s^{(i)}_{3}\cos \phi -s^{(i)}_{2}\sin \phi )+s^{(i)}_{1}\cos \theta \nonumber \\
s_{2}^{(i+1)} & = & -\frac{\sin \theta }{\sqrt{1-\left( s^{(i)}_{3}\right) ^{2}}}(s_{2}^{(i)}s_{3}^{(i)}\cos \phi +s_{1}^{(i)}\sin \phi )+s_{2}^{(i)}\cos \theta \nonumber \\
s^{(i+1)}_{3} & = & \sqrt{1-\left( s_{3}^{(i)}\right) ^{2}}\sin \theta \cos \phi +s_{3}^{(i)}\cos \theta .\label{eq:2}
\end{eqnarray}
 Here \( s^{(i)}_{\alpha } \) etc are stated in the laboratory coordinate
system.

\begin{figure}
{\centering \includegraphics{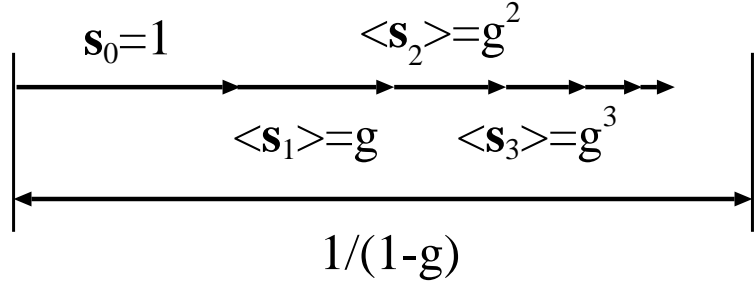} \par}

\caption{The average photon propagation direction (vector) decreases as \protect\( g^{n}\protect \)
where \protect\( g\protect \) is the anisotropy factor and \protect\( n\protect \)
is the number of scattering events.}

\label{fig:photon-direction}
\end{figure}

The ensemble average of the propagation direction over all possible
realizations of \( (\theta ,\phi ) \) and then over all possible
\( \mathbf{s}^{(i)} \) in Eq.~(\ref{eq:2}) turns out to be \( \left\langle \mathbf{s}^{(i+1)}\right\rangle =\left\langle \mathbf{s}^{(i)}\right\rangle \left\langle \cos \theta \right\rangle  \)
because \( \theta  \) and \( \phi  \) are independent and \( \left\langle \cos \phi \right\rangle =\left\langle \sin \phi \right\rangle =0 \).
By recursion, \begin{equation}
\label{eq:3}
\left\langle \mathbf{s}^{(n)}\right\rangle =\left\langle \mathbf{s}^{(0)}\right\rangle \left\langle \cos \theta \right\rangle ^{n}=\hat{z}g^{n}=\hat{z}(1-g_{1})^{n}
\end{equation}
 where \( g=\left\langle \cos \theta \right\rangle =1-g_{1} \) is
the anisotropy factor {[}see Fig.~\ref{fig:photon-direction}{]}.

Using Eq.~(\ref{eq:2}) and recognizing the symmetry obeyed by the
\( x \), \( y \) and \( z \) components of \( \mathbf{s}^{(i)} \),
the correlations between the propagation directions are given by%
\footnote{For example, squaring the third equation in Eq.~(\ref{eq:2}) and
then taking an ensemble average, yields \( \left\langle \left( s^{(i+1)}_{3}\right) ^{2}\right\rangle =\left\langle 1-\left( s_{3}^{(i)}\right) ^{2}\right\rangle \left\langle \sin ^{2}\theta \right\rangle /2+\frac{1}{2}\left\langle \left( s_{3}^{(i)}\right) ^{2}\right\rangle /2=g_{2}/3+(1-g_{2})\left\langle \left( s_{3}^{(i)}\right) ^{2}\right\rangle	 \)
as \( \left\langle \sin ^{2}\phi \right\rangle =\left\langle \cos ^{2}\phi \right\rangle =\frac{1}{2} \).
Similar equalities are obtained for \( x \) and \( y \) components
as the label is rotated.
} \begin{equation}
\label{eq:4}
\left\langle s^{(i+1)}_{\beta }s^{(i+1)}_{\alpha }\right\rangle =\frac{g_{2}}{3}\delta _{\alpha \beta }+(1-g_{2})\left\langle s^{(i)}_{\beta }s^{(i)}_{\alpha }\right\rangle
\end{equation}
where \( g_{2}=\frac{3}{2}\left\langle \sin ^{2}\theta \right\rangle  \).
On the other hand, the correlation between \( s^{(j)}_{\beta } \)
and \( s^{(i)}_{\alpha } \) (\( j>i \)) can be reduced to a correlation
of the form of Eq.~(\ref{eq:4}) due to the following observation\begin{eqnarray}
\left\langle s_{\beta }^{(j)}s_{\alpha }^{(i)}\right\rangle	& = & \int d\mathbf{s}^{(j)}d\mathbf{s}^{(i)}s_{\beta }^{(j)}p(\mathbf{s}^{(j)}|\mathbf{s}^{(i)})s_{\alpha }^{(i)}p(\mathbf{s}^{(i)}|\mathbf{s}^{(0)})\nonumber \\
 & = & \int d\mathbf{s}^{(j-1)}d\mathbf{s}^{(i)}\left[ \int d\mathbf{s}^{(j)}s_{\beta }^{(j)}p(\mathbf{s}^{(j)}|\mathbf{s}^{(j-1)})\right] p(\mathbf{s}^{(j-1)}|\mathbf{s}^{(i)})s_{\alpha }^{(i)}p(\mathbf{s}^{(i)}|\mathbf{s}^{(0)})\nonumber \\
 & = & \int d\mathbf{s}^{(j-1)}d\mathbf{s}^{(i)}gs^{(j-1)}_{\beta }p(\mathbf{s}^{(j-1)}|\mathbf{s}^{(i)})s_{\alpha }^{(i)}p(\mathbf{s}^{(i)}|\mathbf{s}^{(0)})\nonumber \\
 & = & g\left\langle s_{\beta }^{(j-1)}s_{\alpha }^{(i)}\right\rangle \label{eq:5}
\end{eqnarray}
 where \( p(\mathbf{s}^{(j)}|\mathbf{s}^{(i)}) \) means the conditional
probability that a photon jumps from \( \mathbf{s}^{(i)} \) to \( \mathbf{s}^{(j)} \).
Eq.~(\ref{eq:5}) is a result of the Markov property of the process
\( p(\mathbf{s}^{(j)}|\mathbf{s}^{(i)})=\int d\mathbf{s}^{(j-1)}p(\mathbf{s}^{(j)}|\mathbf{s}^{(j-1)})p(\mathbf{s}^{(j-1)}|\mathbf{s}^{(i)}) \)
and the fact \( \int d\mathbf{s}^{(j)}s_{\beta }^{(j)}p(\mathbf{s}^{(j)}|\mathbf{s}^{(j-1)})=gs_{\beta }^{(j-1)} \)
from Eq.~(\ref{eq:2}). Combining Eqs.~(\ref{eq:4}) and (\ref{eq:5}),
and using the initial condition of \( \mathbf{s}^{(0)} \), we conclude\begin{equation}
\label{eq:6}
\left\langle s^{(j)}_{\beta }s^{(i)}_{\alpha }\right\rangle =\frac{\delta _{\alpha \beta }}{3}g^{j-i}\left[ 1+f_{\alpha }(1-g_{2})^{i}\right] ,\quad j\geq i
\end{equation}
 where constants \( f_{1}=f_{2}=-1 \) and \( f_{3}=2 \). Here we
see that the auto-correlation of the \( x \), \( y \), or \( z \)
component of the photon propagation direction approaches \( 1/3 \),
i.e., scattering uniformly in all directions, after a sufficient large
number of scattering (\( \alpha =\beta  \) and \( j=i\to \infty	 \)
), and the cross-correlation between them is zero (\( \alpha \neq \beta  \)).

The connection between the macroscopic physical quantities about the
photon distribution and the microscopic statistics of the photon propagation
direction is made by the probability \( p_{n}(t) \) that the photon
has taken exactly \( n \) scattering events before time \( t \)
(the \( (n+1) \)-th event comes at \( t \)). We claim \( p_{n}(t) \)
obeys the generalized Poisson distribution%
\footnote{Note added in review: The generalized Poisson distribution
  Eq.~(\ref{eq:7}) was previously proven in Lihong Wang and
  S. L. Jacques, Phys. Med. Biol. 39, 2349 (1994).
}
\begin{equation}
\label{eq:7}
p_{n}(t)=\frac{(\mu _{s}t)^{n}\exp (-t)}{n!}=\frac{(\mu _{s}t)^{n}\exp (-\mu _{s}t)}{n!}\exp (-\mu _{a}t)
\end{equation}
 which is the Poisson distribution of times of scattering with the
expected rate occurrence of \( \mu _{s}^{-1} \) multiplied by an
exponential decay factor due to absorption. Here we have used \( \mu _{T}^{-1}=1 \)
as the unit of length. This form of \( p_{n}(t) \) can be easily
verified by recognizing first that \( p_{0}(t)=\exp (-t) \) equals
the probability that the photon experiences no events within time
\( t \) (and the first event occurs at \( t \)); and second that
the probability \( p_{n+1}(t) \) is given by \begin{equation}
p_{n+1}(t)=\int ^{t}_{0}p_{n}(t-t')\frac{\mu _{s}}{\mu _{T}}p_{0}(t')dt'=\int ^{t}_{0}\frac{\left[ \mu _{s}(t-t')\right] ^{n}\exp \left[ -(t-t')\right] }{n!}\frac{\mu _{s}}{\mu _{T}}\exp (-t')dt'=\frac{(\mu _{s}t)^{n+1}\exp (-t)}{(n+1)!},
\end{equation}
 in which the first event occurred at \( t' \) is scattering and
followed by \( n \) scattering events up to but not including time
\( t \), that confirms Eq.~(\ref{eq:7}) at \( n+1 \) if Eq.~(\ref{eq:7})
is valid at \( n \). The total probability of finding a photon at
time \( t \)\begin{equation}
\label{eq:9}
\sum _{n=0}^{\infty }p_{n}(t)=\exp (-\mu _{a}t)
\end{equation}
 decreases with time due to the annihilation of photons due to absorption.

The average propagation direction \( \left\langle \mathbf{s}(t)\right\rangle  \)
at time \( t \) is then, \begin{equation}
\label{eq:10}
\left\langle \mathbf{s}(t)\right\rangle =\frac{\sum ^{\infty }_{n=0}\left\langle \mathbf{s}_{n}\right\rangle p_{n}(t)}{\sum ^{\infty }_{n=0}p_{n}(t)}.
\end{equation}
 Plug Eqs.~(\ref{eq:3}) and (\ref{eq:7}) into Eq.~(\ref{eq:10}),
we obtain\begin{equation}
\left\langle \mathbf{s}(t)\right\rangle =\hat{z}\exp (-\mu _{s}g_{1}t)=\hat{z}\exp (-t/l_{t}).
\end{equation}
 Here \( l_{t}=\mu _{s}^{-1}/(1-g) \) is usually called the transport
mean free path which is the randomization distance of the photon propagation
direction.

\begin{figure}
{\centering \resizebox*{0.75\columnwidth}{!}{\rotatebox{-90}{\includegraphics{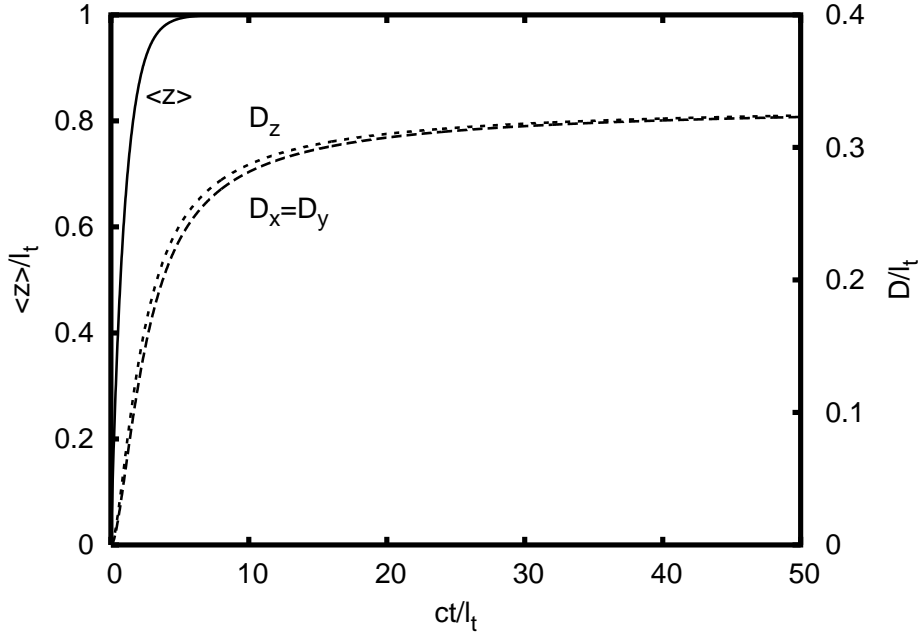}}} \par}

\caption{The center of a photon cloud approaches \protect\( l_{t}\protect \)
along the incident direction and the diffusion coefficient approaches
\protect\( l_{t}/3\protect \) with increase of time.}

\label{fig:center-diffusion}
\end{figure}

The first moment of the photon density with respect to position is
thus\begin{equation}
\left\langle \mathbf{x}(t)\right\rangle =\int ^{t}_{0}\left\langle \mathbf{s}(\tau )\right\rangle d\tau =\hat{z}l_{t}\left[ 1-\exp (-t/l_{t})\right] ,
\end{equation}
 revealing that the center of the photon cloud moves along the incident
direction for one transport mean free path \( l_{t} \) before it
stops {[}see Fig.~\ref{fig:center-diffusion}{]}.

The second moment of the photon density is calculated as follows.
Assume \( p(\mathbf{s}_{2},t_{2}|\mathbf{s}_{1},t_{1}) \) is the
conditional probability that a photon jumps from a propagation direction
\( \mathbf{s}_{1} \) at time \( t_{1} \) to a propagation direction
\( \mathbf{s}_{2} \) at time \( t_{2} \) (\( t_{2}\geq t_{1}\geq 0) \),
the conditional correlation of the photon propagation direction subject
to the initial condition is given by\begin{equation}
\label{eq:13}
\left\langle s_{\beta }(t_{2})s_{\alpha }(t_{1})\right\rangle =\frac{\int d\mathbf{s}_{2}d\mathbf{s}_{1}s_{2\beta }s_{1\alpha }p(\mathbf{s}_{2},t_{2}|\mathbf{s}_{1},t_{1})p(\mathbf{s}_{1},t_{1}|\mathbf{s}_{0},t_{0})}{\int d\mathbf{s}_{2}d\mathbf{s}_{1}p(\mathbf{s}_{2},t_{2}|\mathbf{s}_{1},t_{1})p(\mathbf{s}_{1},t_{1}|\mathbf{s}_{0},t_{0})}.
\end{equation}
 Denote the number of scattering events encountered by the photon
at states \( (\mathbf{s}_{1},t_{1}) \) and \( (\mathbf{s}_{2},t_{2}) \)
as \( n_{1} \) and \( n_{2} \) respectively. Here \( n_{2}\geq n_{1} \)
since the photon jumps from \( (\mathbf{s}_{1},t_{1}) \) to \( (\mathbf{s}_{2},t_{2}) \).
Eq.~(\ref{eq:13}) can be rewritten as\begin{equation}
\label{eq:14}
\left\langle s_{\beta }(t_{2})s_{\alpha }(t_{1})\right\rangle =\frac{\sum _{n_{2}\geq n_{1}}\left\langle s_{\beta }^{(n_{2})}s^{(n_{1})}_{\alpha }\right\rangle p_{n_{2}-n_{1}}(t_{2}-t_{1})p_{n_{1}}(t_{1})}{\sum _{n_{2}\geq n_{1}}p_{n_{2}-n_{1}}(t_{2}-t_{1})p_{n_{1}}(t_{1})}.
\end{equation}
 After a straightforward calculation by utilizing Eq.~(\ref{eq:6})
and (\ref{eq:7}), we obtain\begin{equation}
\label{eq:15}
\left\langle s_{\beta }(t_{2})s_{\alpha }(t_{1})\right\rangle =\frac{\delta _{\alpha \beta }}{3}\left[ 1+f_{\alpha }\exp (-\mu _{s}g_{2}t_{1})\right] \exp \left[ \mu _{s}g_{1}(t_{1}-t_{2})\right] .
\end{equation}
 The second moment is then \begin{eqnarray}
\left\langle x_{\alpha }^{2}(t)\right\rangle  & = & 2\int ^{t}_{0}dt_{2}\int ^{t_{2}}_{0}dt_{1}\left\langle s_{\alpha }(t_{2})s_{\alpha }(t_{1})\right\rangle .
\end{eqnarray}
 The diffusion coefficient is obtained from \( \left( \left\langle \mathbf{x}^{2}\right\rangle -\left\langle \mathbf{x}\right\rangle ^{2}\right) /2t \),
i.e.,

\begin{eqnarray}
D_{xx}=D_{yy} & = & \frac{1}{3t}\left\{ \frac{t}{\mu _{s}g_{1}}+\frac{g_{2}(1-\exp (-\mu _{s}g_{1}t))}{\mu ^{2}_{s}g_{1}^{2}(g_{1}-g_{2})}-\frac{1-\exp (-\mu _{s}g_{2}t)}{\mu ^{2}_{s}g_{2}(g_{1}-g_{2})}\right\} \\
D_{zz} & = & \frac{1}{3t}\left\{ \frac{t}{\mu _{s}g_{1}}-\frac{(3g_{1}-g_{2})[1-\exp (-\mu _{s}g_{1}t)]}{\mu ^{2}_{s}g_{1}^{2}(g_{1}-g_{2})}+\frac{2[1-\exp (-\mu _{s}g_{2}t)]}{\mu ^{2}_{s}g_{2}(g_{1}-g_{2})}-\frac{3[1-\exp (-\mu _{s}g_{1}t)]^{2}}{2\mu _{s}^{2}g_{1}^{2}}\right\} .\nonumber
\end{eqnarray}
 after integration. This exact result does not depend on absorption
and agrees with our previous independently calculated work {[}Eq.~(21)
in Ref.~\onlinecite{cai00:_cumul_boltz}{]}.

The general form of the photon distribution depends on all moments
of the distribution. However, after a sufficient large number of scattering
events have taken place, the photon distribution approaches a Gaussian
distribution over space according to the central limit theorem\cite{		  kendall99}.
This asymptotic Gaussian distribution, characterized by its central
position and half width (\( 2Dt \)), is then \begin{equation}
\label{eq:18}
G(\mathbf{x},t)=C(t)\frac{1}{(4\pi D_{zz}t)^{1/2}}\frac{1}{4\pi D_{xx}t}\exp \left[ -\frac{x^{2}+y^{2}}{4D_{xx}t}-\frac{(z-\left\langle z(t)\right\rangle )^{2}}{4D_{zz}t}\right]
\end{equation}
 where the normalizing factor \( C(t)=\exp (-\mu _{a}t) \) owing
to Eq.~(\ref{eq:9}). This provides a {}``proper'' diffusion solution
to radiative transfer, revealing a behavior of light propagation that
photons migrate with a center that advances in time, and with an ellipsoidal
contour that grows and changes shape {[}see Fig.~\ref{fig:center-diffusion}{]}.

It is also worth mentioning that the absorption coefficient only appears
in the generalized Poisson distribution \( p_{n}(t) \) through an
exponential decay factor \( \exp (-\mu _{a}t) \). This exponential
factor will be canceled in the evaluation of the conditional moments
of the photon distribution {[}see Eqs.~(\ref{eq:13}) and (\ref{eq:14}){]}.
Hence, the sole role played by absorption is to annihilate photons
and affects neither the shape of the distribution function nor the
diffusion coefficient.\cite{	  durduran.ea97:_does,	  cai02:_diffus}

The results, except for the Gaussian photon distribution Eq.~(\ref{eq:18}),
are exact under the sole assumption of a Markov random process of
photon migration. The deviation from a Poisson distribution of scattering
or absorption events can be dealt with by modifying \( p_{n}(t) \).
The Markov random process is usually a good description of scattering
due to short-range forces such as photon migration in turbid media.
In situations where interference of light is appreciable, the phase
of photon, which depends on its full past history, must be considered
and this is non-Markovian. One well-known example is weak localization
of light.\cite{wolf.ea85:_weak} Non-Markov processes may also
occur in scattering involving long-range forces such as Coulomb interaction
between charged particles in which the many-body effect can not be
ignored. However, the idea presented here may still be helpful.

In summary, the macroscopic physical property of photon migration
in turbid medium has its root in the microscopic statistics of photon
propagation direction in direction space which is solely determined
by the phase function. A generalized Poisson distribution function
determined by the scattering and absorption coefficients of the medium
serves as a bridge to connect the microscopic statistics to the macroscopic
property. This provides us a clear and comprehensive physical picture
of photon migration in turbid medium.

This work is supported in part by DOE, NASA IRA, DARPA and PSC CUNY.

\end{document}